# Isotropic non-local Gilbert damping driven by spin currents in epitaxial Pd/Fe/MgO(001) films

Yan Li[1,2], Yang Li[1,2], Qian Liu[3], Zhe Yuan[3], Wei He[1], Hao-Liang Liu[1], Ke Xia[3], Wei Yu[1], Xiang-Qun Zhang[1], and Zhao-Hua Cheng[1,2, *]

[1]*State Key Laboratory of Magnetism and Beijing National Laboratory for Condensed Matter Physics, Institute of Physics, Chinese Academy of Sciences, Beijing 100190, China*

[2]*School of Physical Sciences, University of Chinese Academy of Sciences, Beijing 100049, China*

[3]*The Center for Advanced Quantum Studies and Department of Physics, Beijing Normal University, 100875 China*

**ABSTRACT**

Although both theoretical predications and experimental observations demonstrated that the damping factor is anisotropic at ferromagnet/semiconductor interface with robust interfacial spin-orbit coupling, it is not well understood whether non-local Gilbert damping driven by spin currents in heavy metal/ferromagnetic metal (HM/FM) bilayers is anisotropic or not. Here, we investigated the in-plane angular- and frequency- dependence of magnetic relaxation of epitaxial Fe/MgO(001) films with different capping layers of Pd and Cu. After disentangling the parasitic contributions, such as two-magnon scattering (TMS), mosaicity, and field-dragging effect, we unambiguously observed that both local and non-local Gilbert damping are isotropic in Fe(001) plane, suggesting that the pure spin currents absorption is independent of Fe magnetization orientation in the epitaxial Pd/Fe heterostructure. First principles calculation reveals that the effective spin mixing conductance of

Pd/Fe interface is nearly invariant for different magnetization directions in good agreement with the experimental observations. These results offer a valuable insight into the transmission and absorption of pure spin currents, and facilitate us to utilize next-generation spintronic devices.

*Corresponding author

E-mail: zhcheng@iphy.ac.cn

## I. INTRODUCTION

The rapid development of spintronic devices inquires deeper understanding of the magnetization relaxation mechanism[1-3]. The Gilbert damping factor, one of key parameters in spin dynamics, characterizes the energy transfer from the spin subsystem to the lattice and governs the magnetization switching time and the critical current density in spin transfer torque devices[4-6]. Since the shape of Fermi surface depends on the orientation of the magnetization direction due to the spin-orbit interaction, an anisotropic Gilbert damping is expected in single crystal ultrathin films[7-10]. Chen et al. discovered an anisotropic damping in the Fe/GaAs(001) ultrathin films where an robust interfacial spin-orbit field exists, due to GaAs substrate. The magnitude of damping anisotropy, however, decreases with increasing Fe thickness, and disappears when the Fe thickness is larger than 1.9 nm[11-13].

Besides intrinsic Gilbert damping in ferromagnetic materials (FM), spin currents sink into heavy metals (HM) or other magnetic layers importing non-local Gilbert damping in HM/FM bilayers or spin valve structure according to spin pumping model[14-16]. Although anisotropic magnetization relaxation in ferromagnetic multilayers was observed, it is debated whether the absorption of pure spin currents is anisotropic or isotropic in ferromagnetic multilayers[17-21]. This is because the frequency- and angular-dependent ferromagnetic resonance (FMR) linewidth results are often contaminated by parasitic contributions, such as two-magnon scattering (TMS), mosaicity, and field-dragging effect. Li et al. found that nearly isotropic absorption of pure spin current in Co in $Py_{1-x}Cu_x$/Cu(5 nm)/Co(5 nm) trilayers using spin pumping technique[22]. Meanwhile, Baker et al. found an anisotropic absorption of pure spin currents in $Co_{50}Fe_{50}$/Cr/$Ni_{81}Fe_{19}$ spin valves with variable Cr thickness, while the anisotropy is suppressed above the spin diffusion length[23]. Here, we investigated spin pumping and clarified the dependence of diverse magnetic relaxations on Fe magnetic orientation using Vector Network Analyzer ferromagnetic resonance (VNA-FMR) of epitaxial Fe/MgO(001) films capped by Pd and Cu layers. Simple FM/HM bilayers would be a more convincing candidate to explore the non-local relaxation mechanism. Excluding the misleading dragging effect and the deceitful extrinsic terms, we unambiguously observed that both local and non-local Gilbert damping are isotropic in Fe(001) plane. The isotropic non-local Gilbert damping suggests that the pure spin currents absorption is independent of Fe magnetization orientation, which is supported by the first principles calculation.

## II. EXPERIMENTS

Samples were prepared in molecular beam epitaxy chambers with a basic pressure $2\times10^{-10}$ mbar[24]. Prior to deposition, MgO(001) substrate was annealed at 700 °C for 2 hours, and then 6 nm Fe film was deposited on a MgO(001) substrate using electron-beam gun, and finally 5 nm Pd was covered on Fe films. The crystalline quality and epitaxial relationship was confirmed by high-resolution transmission electron microscopy (HRTEM), as shown in Fig. 1(a) and (b). It has been revealed that the films were grown with the epitaxial relationship Pd(001)<110>||Fe(001)<100>||MgO(001)<110> (see the inset of Fig. 1(b)). For comparison, Cu(3.5 nm)/Fe(6 nm)/MgO(001) sample was also prepared. In-plane

VNA-FMR measurements were performed by facing the sample down on employing a co-planar waveguide (CPW) and recording the transmission coefficient $S_{21}$[25-27]. All depositions and measurements were performed at room temperature.

### III. RESULTS AND DISCUSSION

Fig. 2(a) shows schematically the stacked sample and the measured configuration. The representative FMR spectra at fixed frequency 13.4 GHz and various magnetic field angles $\varphi_H$ are illustrated in Fig. 2(b). The FMR signal (the transmission parameter $S_{21}$) is a superposition of symmetric and antisymmetric Lorentzian functions. The following equation could be used to extract the resonance field $H_r$ and the resonance linewidth $\Delta H$:

$$\mathrm{Re}\, S_{21}(H) = S_0 + L\frac{(\Delta H/2)^2}{(H-H_r)^2+(\Delta H/2)^2} - D\frac{(\Delta H/2)(H-H_r)}{(H-H_r)^2+(\Delta H/2)^2}. \tag{1}$$

Here, Re $S_{21}$, $S_0$, $H$, $L$ and $D$ are the real part of transmission parameter, the offset, the external magnetic field, the symmetric and antisymmetric magnitude, respectively[25-27]. The resonance frequency $f$ is given by Kittel formula[28]

$$f = \frac{\gamma\mu_0}{2\pi}\sqrt{H_a^R H_b^R} \tag{2}$$

with $H_a^R = H_r \cos(\varphi_M - \varphi_H) + H_d + H_4(3+\cos 4\varphi_M)/4 - H_2 \sin^2(\varphi_M - 45°)$,

$H_b^R = H_r \cos(\varphi_M - \varphi_H) + H_4 \cos 4\varphi_M - H_2 \sin 2\varphi_M$ and $H_d = M_s - \frac{2K_{out}}{\mu_0 M_s}$. Here, $\gamma$ and $\mu_0$ are the gyromagnetic ratio and the vacuum permeability. $H$, $H_2$, $H_4$ and $M_s$ are the applied magnetic field, the uniaxial and four-fold magnetic anisotropy fields and saturation magnetization, respectively. $K_{out}$ is the out-of-plane uniaxial magnetic anisotropy constant. The equilibrium azimuthal angle of magnetization $\varphi_M$ is

determined by the following equation:

$$H_r \sin(\varphi_M - \varphi_H) + (H_4/4)\sin 4\varphi_M + (H_2/2)\cos 2\varphi_M = 0. \tag{3}$$

The angular dependent FMR measurements were performed by rotating the samples in plane while sweeping the applied magnetic field. At a fixed frequency of 13.4 GHz, the angular dependence of $H_r$ can be derived from Eq. (2) and plotted in Fig. 2(c) and 2(d) for Fe/MgO(001) samples capped by Pd and Cu, respectively. It can be seen clearly that the angular dependence of $H_r$ demonstrates a four-fold symmetry and the values of $H_2 = 0$ Oe, $H_4 = 625$ Oe and $\mu_0 H_d = 2.0$ T for Pd/Fe/MgO(001) and $H_2 = 0$ Oe, $H_4 = 625$ Oe and $\mu_0 H_d = 1.9$ T for Cu/Fe/MgO(001), respectively. Comparing to the sample with Cu capping layer, Pd/Fe interface modifies the out-of-plane uniaxial magnetic anisotropy, and has a negligible contribution to the in-plane uniaxial magnetic anisotropy.

In contrast to the four-fold symmetry of $H_r$, the angular dependence of $\Delta H$ for the samples with Pd and Cu capping layers indicates to be superposition of four-fold and quasi-eight-fold contributions, as shown in Fig. 3(a) and 3(b), respectively. In fact, the quasi-eight-fold broadening also represents a four-fold symmetry with multiple extreme value points. In the case of the sample with Pd capping layer, $\Delta H$ exhibits two peaks around the hard magnetization directions Fe<110>, and the values of $\Delta H$ for Fe<100> and Fe<110> directions are almost the same (58 Oe). On the other hand, a larger difference in the magnitude of $\Delta H$ was observed along these two directions of Cu/Fe/MgO(001) sample, i.e. 71 Oe and 49 Oe for Fe<100> and Fe<110> axes, respectively.

In order to understand the mechanism of anisotropic magnetic relaxation, we must take both intrinsic and extrinsic contributions into account[29-34]. $\Delta H$ is followed by the expression[32]:

$$\Delta H = \Delta H_{TMS} + \Delta H_{mosaicity} + \Delta H_{Gilbert\_dragging} . \tag{4}$$

The first term denotes TMS, representing that a uniform prerecession magnon ($k=0$) is scattered into a degenerate magnon ($k \neq 0$) due to imperfect crystal structure. Therefore, the contribution of TMS to the linewidth relies heavily on the symmetrical distribution of defects and manifests anisotropic feature accordingly. The second term describes the mosaicity contribution in a film plane, which is caused by a slightly spread of magnetic parameters on a very large scale. The last term $\Delta H_{Gilbert\_dragging}$ is the Gilbert damping contribution with field-dragging.

In the case of Fe/MgO(001) epitaxial film, the contribution of TMS to FMR linewidth composes of numerous two-fold and four-fold TMS channels[31-34],

$$\Delta H_{TMS} = \sum_j \Gamma_{twofold}^{j,\max} \cos^4(\varphi_M - \varphi_{twofold}^{j,\max}) + \sum_j \Gamma_{fourfold}^{j,\max} \cos^2 2(\varphi_M - \varphi_{fourfold}^{j,\max}) . \tag{5}$$

Here, $\varphi_{twofold}^{j,\max}$ and $\varphi_{fourfold}^{j,\max}$ represent angle of the maximum scattering rate in two-fold and four-fold scatterings along the direction *j*. However, the same values of $\Delta H$ between Fe<100> and Fe<110> directions suggest that the TMS can be neglected in Pd/Fe/MgO(001) epitaxial film. On the other hand, the larger difference in the magnitude of $\Delta H$ was observed along these two directions, suggesting that either significant TMS contribution or anisotropic Gilbert damping exists in Cu/Fe/MgO(001) sample[13, 32, 33].

The angular dependence of mosaicity contribution can be described as[32, 34]

$$\Delta H_{mosaicity} = \left|\frac{\partial H_r}{\partial \varphi_H}\right| \Delta \varphi_H, \tag{6}$$

where $\Delta \varphi_H$ represents an in plane variation of mosaicity. $\Delta H_{mosaicity} = 0$ Oe should be hold along easy magnetization directions and hard directions where $\left|\frac{\partial H_r}{\partial \varphi_H}\right| = 0$.

Due to magnetocrystalline anisotropy, magnetization would not always align at the direction of the applied field when the field is weaker than the saturation field. We evaluate the field-dragging effect during rotation of the sample or frequency-swept based on the numerical calculation using Eq. (3). Fig. 4(a) shows $\varphi_H$ dependence on $\varphi_H$ at 13.4 GHz. The relation reveals a conspicuous dragging effect with a four-fold symmetry. At $\varphi_H = 25°$, $|\varphi_H - \varphi_M|$ is as high as $12°$. Fig. 4(b) shows $\varphi_M$ dependence on $f$ at various $\varphi_H$. When the magnetic field is applied along Fe<100> or Fe<110> directions, the magnetization is always aligned along the applied magnetic field. However, there is a conspicuous angle between the magnetization and the magnetic field with the field along intermediate axis. Owing to the angle between magnetization and applied field, $\Delta H$ corresponding to Gilbert contribution with the field-dragging could be disclosed according to the following equations[12, 13]

$$\Delta H_{Gilbert\_dragging} = \Delta[\text{Im}(\chi)] \tag{7}$$

and $\text{Im}(\chi) = \dfrac{\alpha_{eff} \sqrt{H_a^R H_b^R}[H_a H_a + H_a^R H_b^R] M_s}{(H_a H_b - H_a^R H_b^R)^2 + \alpha_{eff}^2 H_a^R H_b^R (H_a + H_b)^2}$, (8)

where $H_a$ and $H_b$ are $H_a^R$ and $H_b^R$ in non-resonance condition. The effective parameter $\alpha_{eff}$ consists of the intrinsic Gilbert damping and the non-local one driven by spin currents.

Generally, $\alpha_{eff}$ was obtained by the slope of the linear dependence of $\Delta H$ on

frequency $f$ along the directions without field-dragging [28]:

$$\Delta H = \frac{4\pi\alpha_{eff} f}{\mu_0 \gamma} + \Delta H_0, \quad (9)$$

where $\Delta H_0$ is inhomogeneous non-Gilbert linewidth at zero-frequency[25-27]. Fig. 5 shows $\Delta H$ dependence on frequency at various $\varphi_H$. Obviously, $\Delta H$ versus $f$ can be fitted linearly with $\alpha_{Pd/Fe} = 6.0\times 10^{-3}$ and $\alpha_{Cu/Fe} = 4.2\times 10^{-3}$ for magnetic field along easy axes Fe<100> or hard axes Fe<110> of the samples with Pd and Cu capping layers, respectively, indicating isotropic damping (Fig. 5(f) and 5(j)). By using the aforementioned isotropic damping factors, the contributions of TMS, mosaicity, and field-dragging effect are separated from the angular dependence of $\Delta H$ (Fig. 3 a-b). Table I summarizes the fitted parameters in the two samples. Compared with Cu/Fe sample, one observes a significant reduction of mosaicity broadening and a negligible TMS term in Pd/Fe bilayers. In fact, due to high mobility, the capping layer Cu forms nanocrystallites on Fe film, which causes interfacial defects dependence on the crystallographic axes[35-38]. The interfacial defects will impact a four-fold linewidth broadening due to TMS. In contrast, the excellent epitaxial quality at Pd/Fe interface not only ensures a sharp interfacial structure, but also reduces defect density to decrease TMS contribution. Moreover, the mosaicity contribution, indicating the fluctuation of the magnetic anisotropy field, could be strengthen by the interfacial stacking faults. Consequently, a fully epitaxial structure could significantly decrease the extrinsic contributions, especially TMS and mosaicity terms.

Taking these contributions to magnetization relaxation into account, the frequency dependence of $\Delta H$ at various directions can be well reproduced, as shown in Fig. 5(f)-(j). For other directions rather than Fe<100> and Fe<110>, nonlinear relationship between $\Delta H$ and $f$ are evident and illustrated in Fig. 5(g-i). At $\varphi_H = 20°$, the $\Delta H$ vs $f$ curve brings out a slight bump comparing to the linear ones along hard or easy axes. At $\varphi_H = 27°$, $\Delta H$ has a rapid decrease after $\Delta H$ experiencing an abrupt enhancement. At $\varphi_H = 33°$, $\Delta H$ decreases more sharply after 11 GHz. The nonlinearity can be ascribed to the parasitic contributions, such as TMS, mosaicity, and field-dragging effect. It is virtually impossible to stem from TMS for the distorted curves because a nonlinear linewidth broadening due to TMS increases as frequency increases, and approaches to saturation at high frequency[31]. According to the calculation in Fig. 4(b), there is a huge field-dragging effect except the applied magnetic field $H$ along hard and easy axes. The field-dragging will make $\Delta H$ vs $f$ deviate from the linear relationship. As expected, we could effectively fit the experimental data $\Delta H$ vs $f$ using the following equation in association with the original formulas (7),

$$\Delta H = \Delta[\text{Im}(\chi)] + \Delta H_0. \tag{10}$$

Eq. (10) converges to the Eq. (9) with the applied magnetic field along the directions without field-dragging, i.e. easy axes Fe<100> or hard axes Fe<110>[13].

After distinguishing the contributions of extrinsic terms and field-dragging effect, the Gilbert damping factors $\alpha_{eff}$ along various directions are shown in Fig. 6(a). According to the classical spin pumping model[14], precessional magnetization in FM layer will pump spins into adjacent nonmagnetic metals across interface. Cu with only

*s* conduction band has a smaller spin-flip probability and a larger spin diffusion length than 500 nm[39], therefore, the reference sample Cu/Fe cannot increase the Gilbert damping due to a capping layer Cu. In contrast, Pd-layer with strong spin-orbit coupling has a larger spin-flip probability, the injected spin currents are dissipated in Pd-layer, and enhance the intrinsic Gilbert damping of Fe film. The enhancement of the Gilbert damping allows us to comprehend the non-local relaxation mechanism. Obviously, it can be seen from Fig. 6(a) that there is no strong relation between the non-local Gilbert damping and the magnetization orientation in epitaxial film Pd/Fe. The parameters $\alpha_{Cu/Fe}=4.2\times10^{-3}$ and $\alpha_{Pd/Fe}=6.0\times10^{-3}$ are the Gilbert damping of Pd/Fe and Cu/Fe, respectively. The non-local Gilbert damping could be evaluated using the effective spin mixing conductance $g_{eff}^{\uparrow\downarrow}$ [14]

$$\Delta\alpha=\alpha_{Pd/Fe}-\alpha_{Cu/Fe}=\frac{g\mu_B}{4\pi M_s t_{Fe}}g_{eff}^{\uparrow\downarrow}. \tag{11}$$

The obtained isotropic value $g_{eff}^{\uparrow\downarrow}=1.23\times10^{19}\ m^{-2}$ is comparable to the literatures[40-42].

In order to theoretically investigate the dependence of the non-local Gilbert damping on the magnetization orientation, the first principles calculation was performed to calculate the total Gilbert damping of the Pd/Fe/Pd multilayer on the basis of the scattering theory[43-45]. The electronic structure of the Pd/Fe interface was calculated self-consistently using the surface Green's function technique implemented with the tight-binding linearized muffin-tin orbitals method. Within the atomic sphere approximation, the charge and spin densities and the effective Kohn-Sham potentials were evaluated inside atomic spheres. The total Gilbert damping was then calculated using the scattering theory of magnetization dissipation[45]. We simulated the room temperature via introducing frozen thermal lattice disorder into a 5x5 lateral supercell[43]. The root-mean-squared displacement of the atoms is determined by the Debye model with the Debye temperature 470 K. A 28x28 k-mesh is used to sample the two-dimensional Brillouin zone and five different configurations of disorder have been calculated for each Fe thickness. The total Gilbert damping exhibits a linear dependence on the length of Fe and the intercept of the linear function can be extracted corresponding to the contribution of the spin pumping at the Pd/Fe interface[44]. The interfacial contribution is converted to the effective spin mixing conductance, plotted in Fig. 6(b) as a function of the magnetization orientation. It can be seen that the effective spin mixing conductance across Pd/Fe interface $g_{eff}^{\uparrow\downarrow}=1.29\times10^{19}\ m^{-2}$ is independent of the magnetization direction, and is in very good agreement with the experimental value $1.23\times10^{19}\ m^{-2}$. According to the Elliott-Yafet mechanism in a nonmagnetic metal, spins relax indiscriminately energy and momentum along all orientation in Pd-layer since a cubic metal is expected to possess a weak anisotropy of the Elliott-Yafet parameter[46]. Incidentally, the fitting error will mislead an anisotropic Gilbert damping if ones use Eq. (9) to fit the entire curves $\Delta H$ vs $f$. Besides, an epitaxial magnetic film integrated into a pseudo spin valve could lead to an anisotropic absorption of spin current based on spin transfer torque mechanism since it is demanding to drag magnetization paralleling to the applied field[11].

# IV. CONCLUSIONS

In summary, a non-local Gilbert damping is induced by the spin pumping in Pd/Fe bilayers as spin currents transfer angular momentum into the Pd-layer. Due to strong magnetocrystalline anisotropy, the field-dragging effect makes the linewidth versus frequency deviate from the linear relationship except magnetic field along hard or easy axes. Extrinsic relaxation, such as TMS and mosaicity, relies heavily on magnetization orientation. However, an epitaxial interface could significantly decrease and minimize the extrinsic contributions, especially TMS and mosaicity. It is noteworthy that an isotropic non-local Gilbert damping factor is clarified after ruling out the misleading field-dragging effect and the deceitful extrinsic contributions. Magnetization orientation has a negligible contribution to the non-local Gilbert damping based on both theoretical and experimental results, manifesting that the absorption of pure spin currents across interface Pd(100)[110]/Fe(001)[100] is independent of Fe magnetization orientation. Our works provide deeper insight into the non-local Gilbert damping mechanism.

ACKNOWLEDGMENTS

This work is supported by the National Key Research Program of China (Grant Nos. 2015CB921403, 2016YFA0300701, and 2017YFB0702702), the National Natural Sciences Foundation of China (Grant Nos. 51427801,1187411,51671212, and 11504413) and the Key Research Program of Frontier Sciences, CAS (Grant Nos. QYZDJ-SSW-JSC023, KJZD-SW-M01 and ZDYZ2012-2). The work at Beijing Normal University is partly supported by the National Natural Sciences Foundation of China (Grant Nos. 61774017, 61704018, and 11734004), the Recruitment Program of Global Youth Experts and the Fundamental Research Funds for the Central Universities (Grant No. 2018EYT03).

**FIGURE CAPTIONS**

Fig. 1 **(Color online)** (a) Dark field scanning high-resolution transmission electron microscopy image and (b) selected area electron diffraction pattern of Pd/Fe/MgO(001). The inset of Fig. 1(b) shows a schematic of the epitaxial relationship.

Fig. 2 **(Color online)** (a) A schematic illustration of the stacked sample Pd/Fe/MgO(001). The sample is placed on the CPW for FMR measurement, and could be rotated in plane. (b) Typical real FMR spectra of Pd/Fe at fixed frequency 13.4 GHz at various magnetic field angles $\varphi_H$. Magnetic field angle $\varphi_H$ dependence of the resonance field $H_r$ at a fixed frequency 13.4 GHz for Pd/Fe (c) and Cu/Fe (d). The red curves are fit to Kittel's formula (2). (In order to show clearly the tendency, we show the data at $-45° \leq \varphi_H \leq 225°$, the same below)

Fig. 3 **(Color online)** The measured linewidth $\Delta H$ as a function of $\varphi_H$ at 13.4 GHz for Pd/Fe (a) and Cu/Fe (b). The linewidth $\Delta H$ is superimposed by several terms, such as TMS, mosaicity and Gilbert contribution with field-dragging.

Fig. 4 **(Color online)** Field-dragging effect for Pd/Fe. (a) The green line denotes the equilibrium direction of magnetization as a function of magnetic field angle $\varphi_H$ at 13.4 GHz. The red line indicates the misalignment between the magnetization and the applied magnetic field accordingly. (b) The equilibrium direction of the magnetization in the frequency-swept mode at various $\varphi_H$.

Fig. 5 **(Color online)** Frequency dependence of the resonance field $H_r$ (a-e) and frequency dependence of the resonance linewidth $\Delta H$ (f-j) for Pd/Fe at various $\varphi_H$.

The blue solid squares and curves in (f) and (j) corresponding to frequency dependence of $\Delta H$ at $\varphi_H = 0°$ and $\varphi_H = 45°$ for Cu/Pd.

Fig. 6 **(Color online)** Angular dependent Gilbert damping and first principles calculation. (a) The opened and solid green squares represent the obtained Gilbert damping for Pd/Fe and Cu/Fe films, respectively. The red and blue lines are guide to the eyes. (b) The experimental and calculated spin mixing conductance as a function of the orientation of the equilibrium magnetization.

**Table I** The fitted magnetic anisotropy parameters and magnetic relaxation parameters in Pd/Fe and Cu/Fe films.

**Fig.1**

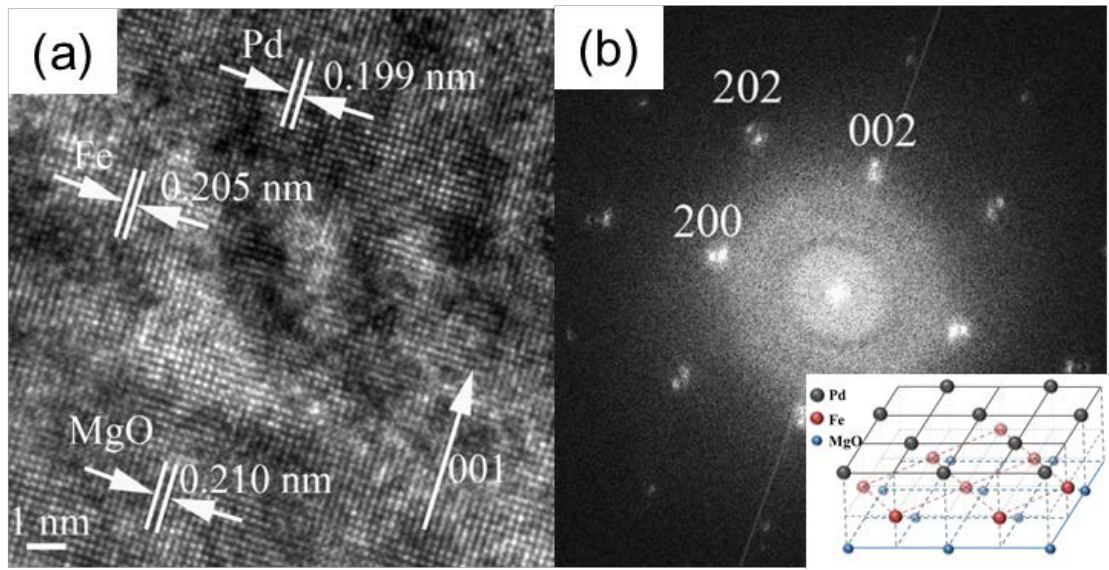

**Fig. 2**

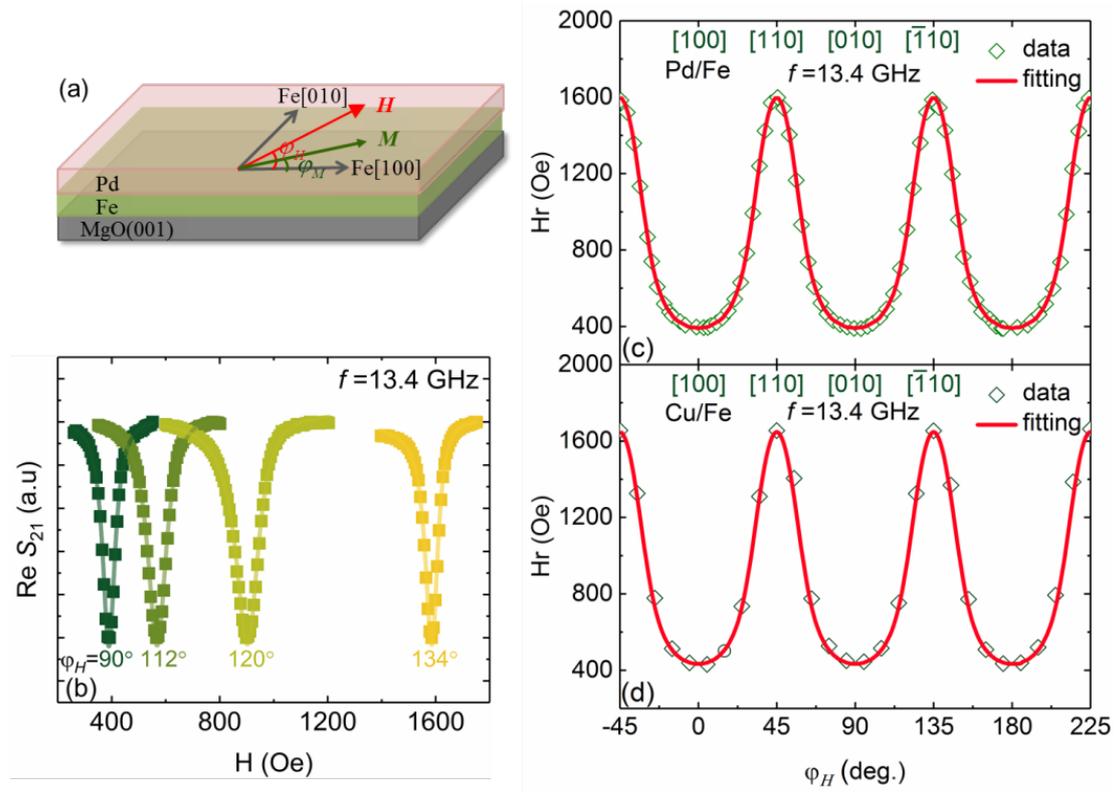

**Fig. 3**

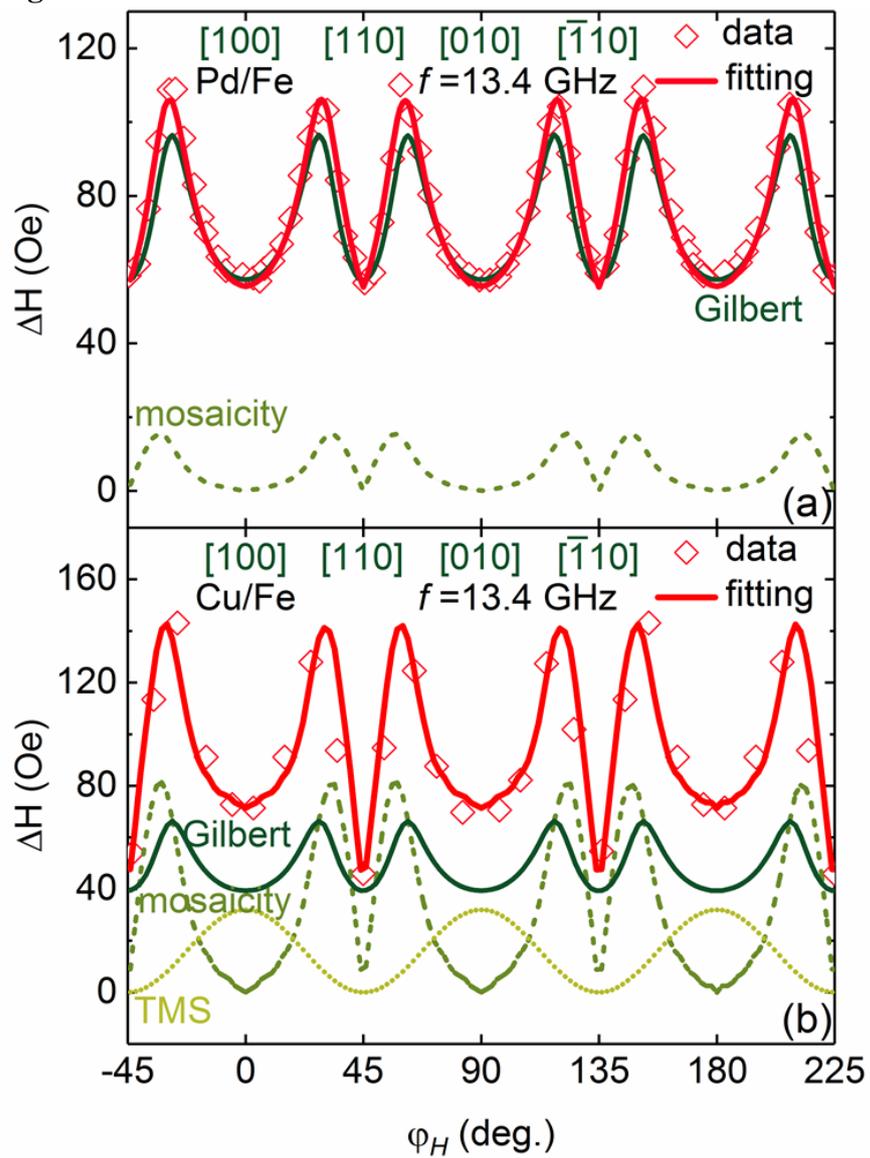

**Fig. 4**

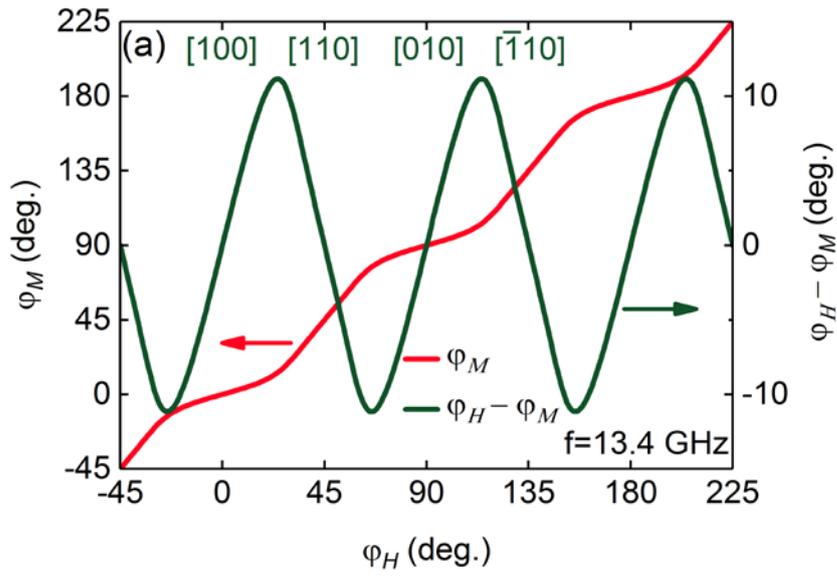

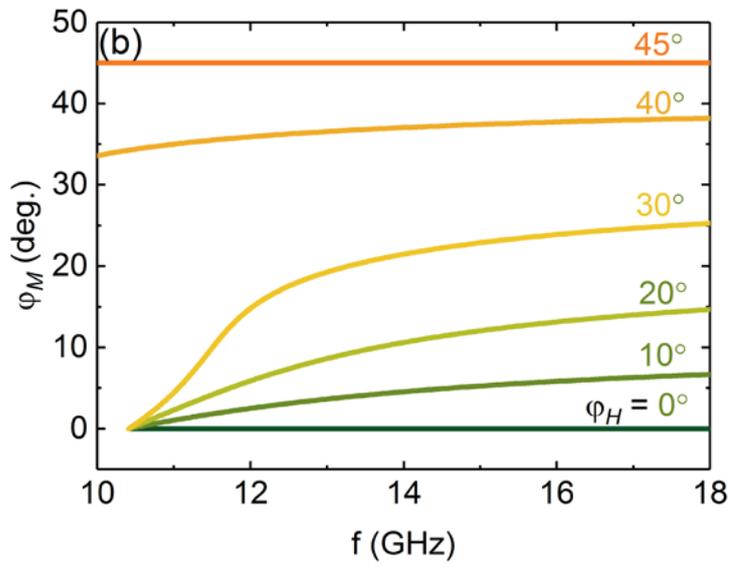

**Fig. 5**

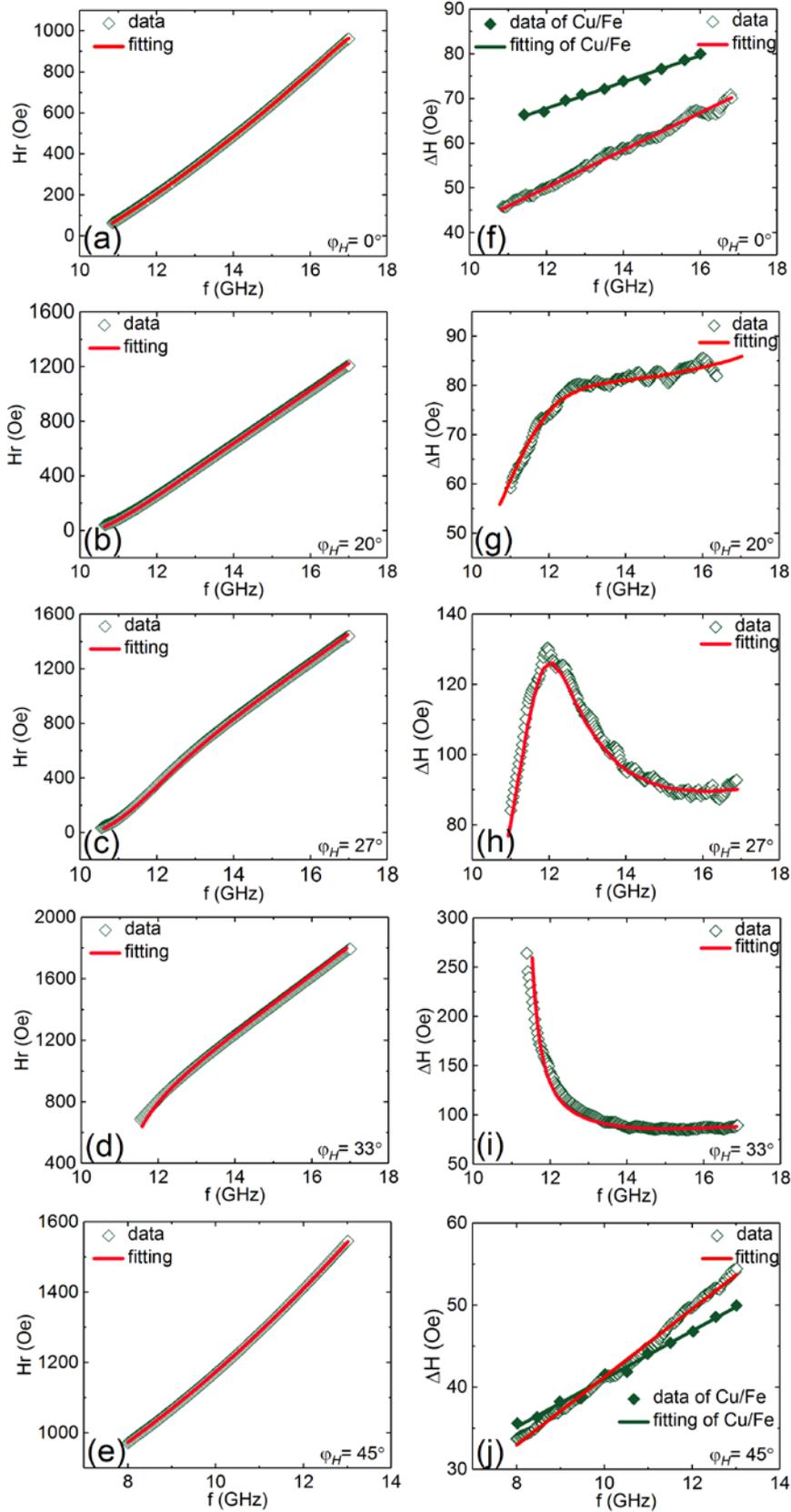

**Fig. 6**

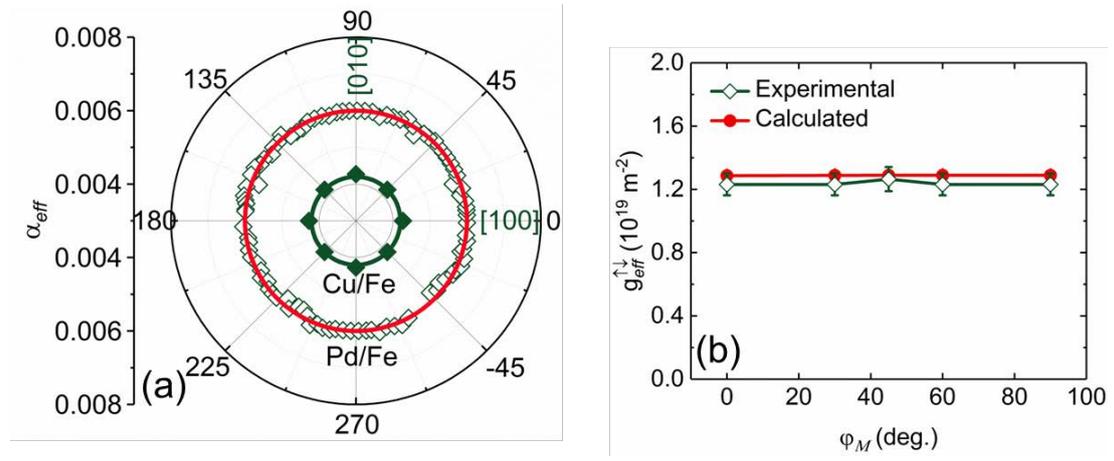

Table I The fitted magnetic anisotropy parameters and magnetic relaxation parameters in Pd/Fe and Cu/Fe films in Fig. 3.

| Sample | $H_4$ (Oe) | $H_2$ (Oe) | $\mu_0 H_d$ (T) | $\alpha_{eff}$ | $\gamma\Gamma_{<100>}$ ($10^7$ Hz) | $\Delta\varphi$ (deg.) |
|---|---|---|---|---|---|---|
| **Pd/Fe** | 625 | 0 | 2.0 | 0.0060 | 0 | 0.23 |
| **Cu/Fe** | 625 | 0 | 1.9 | 0.0042 | 58 | 1.26 |